\documentclass[12pt,a4paper,dvips]{article}
\usepackage{epsfig,wrapfig,times,mathptm} 
\renewcommand{\section}[1]{\vspace{6pt} \noindent\mbox{#1} \newline \noindent}
\renewcommand{\subsection}[1]{\vspace{6pt} \noindent\mbox{\underline{#1}} 
\newline \noindent}
\renewcommand{\subsubsection}[1]{\vspace{6pt} \noindent\mbox{\underline{#1}}
\noindent}

\newfont{\sansb}{cmssbx10}
\newfont{\sans}{cmss10}

\begin{document}
{\small OG 10.3.32 \vspace{-24pt}\\}     
{\center \LARGE   CASA-BLANCA: A LARGE NON-IMAGING CERENKOV DETECTOR
AT CASA-MIA
\vspace{6pt}\\}
M. Cassidy$^1$,
L. F. Fortson$^1$, J. W. Fowler$^1$, C. H. Jui$^2$, D. B. Kieda$^2$, 
E. C. Loh$^2$, R. A. Ong$^1$,
P. Sommers$^2$ \vspace{6pt}\\
{\it $^1$Enrico Fermi Institute, University of Chicago, Chicago, IL 60637\\
$^2$Department of Physics, University of Utah, Salt Lake City, UT 84112 \vspace{-12pt}\\}

{\center ABSTRACT\\}
The lateral distribution of Cerenkov light at ground level records
important information on the development of the air shower which
produces it.  We have constructed a Broad Lateral Non-imaging Cerenkov
Array (BLANCA) to measure this lateral distribution at the CASA-MIA
air shower detector in Dugway, Utah.  Together, the arrays can sample
the lateral distributions of electrons, muons, and Cerenkov light, at
many well-measured distances from the shower core.
We describe the design and calibration of the CASA-BLANCA
experiment and its ability to address the composition of
primary cosmic rays between $3 \times 10^{14}$ and $3 \times 10^{16}$ eV. 
\setlength{\parindent}{1cm}

\begin{wrapfigure}{r}{9cm}
\epsfig{file=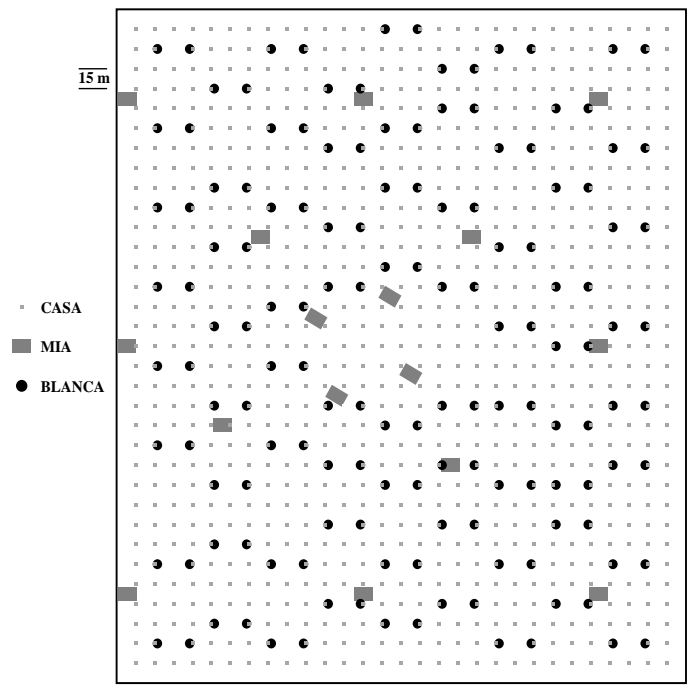,width=9cm,height=9cm}
\caption{\it The CASA-MIA and BLANCA detector arrays.}
\label{map}
\end{wrapfigure}

\section{INTRODUCTION}
Ground-based measurements of the cosmic ray energy spectrum show a
clear break in the vicinity of $3\times 10^{15}$ eV, known as the
spectral ``knee.''  Over a range of several decades lower in energy,
the spectrum falls as $E^{-2.7}$; at higher energies, the power law
index steepens to $\sim~3.15$ (Gaisser et al., 1995).  This change in
slope is not understood.  The 
favored supernova shock wave model for cosmic ray acceleration does
give rise to a natural cut-off in energy, due to the limited duration
of the supernove explosion,
but it fails to explain the spectrum above that cut-off (Drury et al.,
1994).  The
elemental composition of cosmic rays can provide clues to this
problem, because both the processes of shock acceleration and escape
from the galaxy depend on the magnetic rigidity of the particle.  Any
cut-off energy should be higher for particles with greater charge,
hence a heavier composition above the knee.  Speculations of
high energy cosmic rays coming from extra-galactic sources, however,
lead to predictions of progressively lighter composition at higher
energy (Protheroe and Szabo, 1992).  Unfortunately, given the low
cosmic ray flux at high energies, direct measurements of the
composition at and above the knee are not now possible.

Since a ground-based cosmic ray experiment does not directly observe the
parameters of astrophysical interest---cosmic ray energy and nuclear
charge---it must instead measure as many complementary properties of
the air shower as possible.  The lateral distribution of Cerenkov
light about the core of the shower is one such property.  Air shower
simulations show that more than $\sim 150$ m from the shower core, 
the density of Cerenkov light is quite insensitive to the primary cosmic
ray species yet is proportional to the primary's energy (Patterson and
Hillas, 1983).
The light density profile near the core is much more sensitive to the
penetration depth of the shower into the atmosphere, which in turn
correlates closely with the mass of the primary particle: lighter
nuclei penetrate more deeply.  We have constructed the BLANCA array to
sample the Cerenkov light distribution.
Other important shower properties are also measured at the same
site, including the number and distribution of muons and
electromagnetic particles.

\section{THE CASA-MIA INSTRUMENT}
The CASA-MIA experiment, located in Dugway, Utah, U.S.A. ($40.2^\circ$
N, $112.8^\circ$ W), consists of a surface array (CASA) of 1089 scintillation
detector stations and a muon array (MIA) of 1024 scintillation
counters (Borione et al., 1994).  The general arrangement is shown in
Figure~\ref{map}.  Buried three meters underground in
sixteen patches, the muon counters have a total active area of 2500 m$^2$.
The surface array now operates with only 885 stations, enclosing an
area of $2.0 \times 10^5 $~m$^2$.  Each surface station has 1.3~m$^2$ of
scintillator area.   The location of air shower cores is determined by
finding the point of maximum surface particle density, and shower
arrival directions are reconstructed using surface detector timing
information.  For cosmic rays of interest here---{\em i.e.} above 300
TeV---the core uncertainty is $\sim 3$~meters, and the angular
resolution is less than $1^\circ$.  CASA-MIA has been in continuous
operation since January, 1992.

\begin{wrapfigure}{r}{6cm}
\epsfig{file=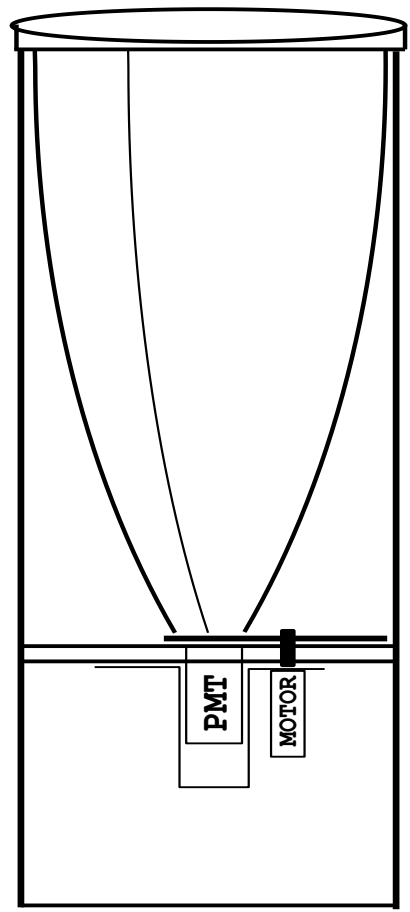,width=6cm,height=12cm}
\caption{\it A BLANCA detector.}
\label{sectionfig}
\end{wrapfigure} 

To the existing CASA-MIA experiment, we have added 144 non-imaging
Cerenkov light detectors.  These detectors are arrayed as show in
Figure~\ref{map}, with a typical spacing of $\sim 35$ m.
The BLANCA array has no trigger mechanism of its own.
Instead, the detector pairs decide locally whether to record data each
time the CASA surface array triggers.  Since the estimated threshold
energy at CASA even for heavy nuclei is below 300 TeV, this method is
fully efficient at capturing the high energy showers sought by BLANCA.

\section{BLANCA CERENKOV DETECTORS}
The BLANCA units each contain a large
Winston cone, concentrating light on a 3-inch photomultiplier tube
(Figure~\ref{sectionfig}).  The cone is oriented vertically and has a
nominal acceptance half-angle of $12.5^\circ$.  It is made of clear
plastic, aluminized on its inner surface through vacuum deposition.
The PMT is placed in a magnetically shielded housing, with
its window as close as possible to the output aperture
of the cone.  However, a rotating shutter is placed between them
to protect the photocathode from sunlight.  At night, a motor can
be commanded remotely to rotate this shutter into its open position.
A two-gain pre-amplifier mounted near the tube housing amplifies the PMT
pulses by factors of 40 and 1.5.  The second (low-gain) channel
alleviates the dynamic range restrictions that would otherwise result
from ADC saturation on the high-gain channel.  When combined, these
two outputs give BLANCA sensitivity to Cerenkov light signals varying
over three orders of magnitude.  Both pre-amp outputs are
connected to a modified CASA station board 15 meters away.  In
this configuration, the four QDC channels on a single board can serve
two BLANCA units, hence the paired geometry seen in Figure~\ref{map}.
The same board controls the shutter operations and the high voltage
supplied to the detector pair.  For protection from the elements, the
BLANCA units fit into PVC enclosures 37 cm in diameter.  An equally
large piece of UV-transmitting glass sealed to the top of each Winston
cone keeps rain and moisture out of the entire assembly.  Finally, an
8-Watt heater wire warms the glass to melt snow and eliminate frost.

\begin{table}
\begin{center}
\begin{tabular}{||c|c||}\hline
Number of detectors & 144 \\ \hline
Average spacing & 35 m \\ \hline 
Overall enclosed area & 201,600 m$^2$ \\ \hline
Cone half-angle & $12.5^\circ$ \\ \hline
 & Length: 61 cm \\
Cone size & Entrance Diam: 36 cm \\
 & Exit diam: 7.6 cm \\ \hline
PMT diameter & 8.3 cm \\ \hline
\end{tabular}
\end{center}
\caption{\it Physical characteristics of BLANCA and its detector units.}
\end{table}

\begin{wrapfigure}{r}{8cm}
\epsfig{file=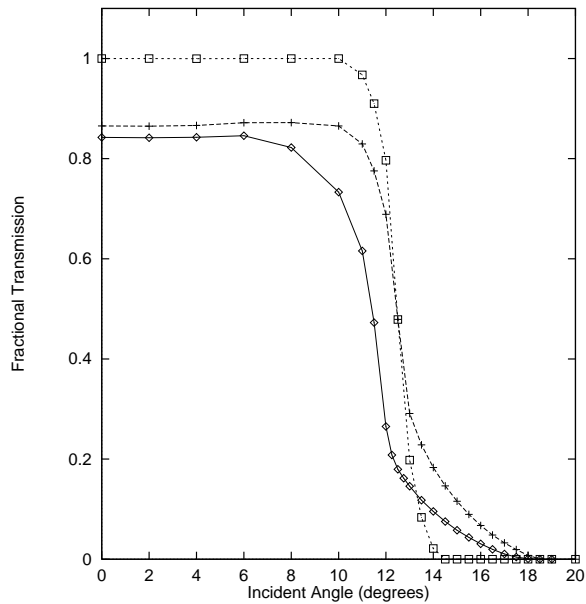,width=8cm,height=8cm}
\caption{\it Fraction of collimated light entering the Winston cone which is
transmitted to the PMT face, as a function of entrance angle.  The
dotted curve represents an ideal $12.5^\circ$ cone; the dashed curve
assumes a 90\% mirror reflectivity and cone truncation at 61~cm
length; the solid curve also adds the effect of the 6~mm gap
between BLANCA cones and the phototubes.}
\label{cone}
\end{wrapfigure} 

The Winston cone used in each BLANCA serves both as a
collimator---substantially reducing night sky background light---and a
concentrator.  These cones were manufactured based on the exact form
of a Winston cone with acceptance half-angle of $12.5^\circ$ and exit
aperture having 90\% the diameter of our PMTs.  Such an optical
element would transmit almost all light arriving inside its
acceptance cone and reflect almost all light outside it (Welford and
Winston, 1989).  However, the BLANCA cone was truncated to only 
61 cm, or 63\% of the ideal length.  This alteration reduces the
collecting area from 970 cm$^2$ to 880 cm$^2$ and also compromises
the collimating effect somewhat.  The necessity of placing protective
shutters between the cones and PMTs effectively reduces the cones'
acceptance angle to roughly $11^\circ$.  Figure~\ref{cone} shows the
transmission of ideal and more realistic cones as a function of
entrance angle, calculated with ray-tracing algorithms.

The 10-stage EMI photomultiplier tubes used in BLANCA were each calibrated
at the University of Utah before installation.  The calibration system
used a continuous beam helium-cadmium laser (325 nm) with constant and
known intensity, measuring the tube anode current at four different
voltages.  We used these results to operate all BLANCA tubes in the
array as nearly as possible at the same gain.  We also had four typical
PMTs tested by Philips to confirm that all had similar cathode
sensitivities between 300 and 500~nm.

The full array of 144 BLANCA detectors was constructed and installed
at the Dugway site between July, 1996 and January, 1997.  From its
completion to mid-May, 1997, CASA-BLANCA has recorded 200 hours of
Cerenkov data.  Using the Akeno measurement of the cosmic ray
spectrum (Gaisser et al., 1995), we estimate that this data sample
contains 4000 air showers above $2 \times 10^{15}$ eV and 170 above
$10^{16}$ eV.  With anticipated running in the winter of 1998, we
hope at least to double this data set.

\section{LIGHT DETECTOR CALIBRATION}
Converting observed ADC values into Cerenkov light density
measurements requires a calibration of the sensitivity of
each BLANCA detector.  We consider the independent performance
of the elements separately to get a first estimate of this
sensitivity.  Combining the optical properties of the glass top and
the cone, the lab results on the PMT gains, and studies of the pre-amp and
digitizing circuit, we estimate that one ADC count corresponds to about
0.1 blue Cerenkov photons per cm$^2$ above the glass detector lids.  BLANCA
should be able to measure Cerenkov photon densities roughly between 1--2
cm$^{-2}$ and 3000 cm$^{-2}$.  Night sky background light and
electronic noise limit the lowest accessible density; ADC saturation
even on the low-gain pre-amplifier output limits measurements of very
high photon densities.

On a nightly basis, the relative sensitivities of the BLANCA stations
can be found from the Cerenkov data themselves. This procedure relies
only on the uniform distribution of air shower cores over the array,
which can be checked using CASA alone.  We alos expect to monitor time
variations of the PMT and amplifier gains using shower
Cerenkov light, under the assumption that the energy spectrum of
Cosmic ray showers is steady from night to night.

With these methods of finding the relative sensitivities between tubes
and between observing periods, the question remains how to
find the absolute response of the array to Cerenkov light.  We have
begun to develop a portable flasher incorporating a blue LED and
collimating optics which can be carried to several BLANCA detectors in
one night.  The LED light pulse output would be determined by flashing
a calibrated PMT.  Also, BLANCA has recorded flashes of light from the
HiRes Laserscope (Bird, 1996)---a nitrogen laser aimed horizontally
150~m above the entire BLANCA array.  Since the amount of light scattered
into a detector depends only on the elevation angle of the laser path
as seen by that detector, these data can provide an external check on
the accuracy of our relative calibrations.  Calculations of the amount
of scattered laser light will also help to establish the absolute
detector sensitivities.

\section{CONCLUSIONS}
The main strength of the CASA-BLANCA experiment is its ability to
measure multiple components of incident air showers.  BLANCA itself,
with 144 separate stations, insures several Cerenkov light density
samples near a shower core.   
By measuring this density and other shower
parameters simultaneously, we will vastly improve our ability to
determine the cosmic ray composition at the knee.

\section{ACKNOWLEDGEMENTS}
We thank the entire CASA-MIA collaboration for countless contributions
to this project. We acknowledge the extensive assistance of
the University of Utah High-Resolution Fly's Eye (HiRes) group
and the staff of the U.S. Army Dugway Proving Ground.  We also thank
D. Bird, J. Meyer, M. Pritchard, K. Riley,  S. Thomas, W. Wingert.
This work is supported by the U.S. National Science
Foundation and the U.S. Department of Energy.  RAO acknowledges the
support of the Grainger Foundation and the Alfred P. Sloan Foundation.

\section{REFERENCES}
\setlength{\parindent}{-5mm}
\begin{list}{}{\topsep 0pt \partopsep 0pt \itemsep 0pt \leftmargin 5mm
\parsep 0pt \itemindent -5mm}
\vspace{-15pt}
\item Bird, D., private communication (1996).
\item Drury, L., et al., {\em Astron. Astrophys.} {\bf 287}, 959 (1994).
\item Gaisser, T. K., et al., in {\em Proc. 1994 Snowmass Summer
Study}, ed. E. W. Kolb and R. D. Peccei, pp. 273--294, World
Scientific, New Jersey (1995).
\item Patterson, J. R., and Hillas, A. M., {\em J. Phys. G:
Nucl. Phys}, {\bf 9}, 1433 (1983).
\item Protheroe, R. J., and Szabo, A. P., {\em Phys. Rev. Lett.} {\bf
69}, 2885 (1992).
\item Welford, W. T. and Winston, R., {\em High Collection Non-imaging
Optics}, Academic Press, San Diego (1989).
\end{list}
\end{document}